\documentclass[useAMS,usenatbib]{mn2e}
\usepackage{graphicx,rotate,url,lscape,times,mathptmx}
\usepackage{color}
\usepackage{latexsym}
\usepackage{amsmath}
\usepackage{multirow}
\newcommand{\etal}{et al. }
\newcommand{\Ubar}{\overline{U}}
\newcommand{\rme}{{\rm e}}
\newcommand{\kb}{k_{\rm B}}
\def\min{{\rm min}}
\def\max{{\rm max}}

\def\rmc{{\rm c}}
\def\rme{{\rm e}}

\def\rmp{{\rm p}}
\def\rmd{{\rm d}}

\def\rmm{{\rm m}}

\def\rmH{{\rm H}}
\def\rmI{{\rm I}}
\def\rmII{{\rm II}}

\def\balpha{\boldsymbol{\alpha}}

\def\be{\begin{equation}}
\def\ee{\end{equation}}
\def\beq{\begin{eqnarray}}
\def\eeq{\end{eqnarray}}
\def\gtrsim{\mathrel{\hbox{\rlap{\hbox{\lower4pt\hbox{$\sim$}}}\hbox{$>$}}}}
\def\lsrsim{\mathrel{\hbox{\rlap{\hbox{\lower4pt\hbox{$\sim$}}}\hbox{$<$}}}}
\def\half{\frac{1}{2}}
\def\3half{\frac{3}{2}}

\def\Ubar{\overline{U}}

\title[H, He-like recombination spectra IV]{H, He-like recombination spectra IV:  
clarification and refinement of methodology for $l$-changing collisions}
\author[N. R. Badnell et al.]
       {\parbox[]{6.0in}
        { N. R. Badnell$^1$,  F. Guzm\'an$^2$\thanks{Present address: Department of Physics and Astronomy,
                                                   University of North Georgia, Dahlonega, GA 30597, USA}, 
        S. Brodie$^1$,  R. J. R. Williams$^3$, P. A. M. van Hoof$^4$, M. Chatzikos$^2$ and G. J. Ferland$^{2}$. \\
        \footnotesize
        $^1$Department of Physics, University of Strathclyde, Glasgow G4 0NG, UK\\
        $^2$Department of Physics and Astronomy, University of Kentucky, Lexington, KY 40506, USA\\
        $^3$Department of Physics and Astronomy, University of Leeds, Leeds LS2 9JT, UK \\
        $^4$Royal Observatory of Belgium, Ringlaan 3, 1180 Brussels, Belgium}
}

\date{%Accepted .
      Received }

\pagerange{\pageref{firstpage}--\pageref{lastpage}}
\pubyear{}

\begin{document}

\maketitle

\label{firstpage}

\begin{abstract}
Precise spectral diagnostic modelling of H~{\sc i} and He~{\sc ii} recombination spectra can constrain 
theoretical models which describe many astrophysical environments.
Simple analytic expressions are of interest for collisional $l$-changing rate coefficients that 
are used by large-scale population modelling codes.
We review, clarify and improve-upon the modified Pengelly \& Seaton formulae of Guzm\'an \etal
We show that the recent poor results for it shown by Vrinceanu \etal are due to their
misinterpretation of its usage. We also detail efficient numerical algorithms which
should enable the full quantum mechanical expression for such rate coefficients to be used 
much more routinely by modelling codes. We illustrate with some collisional-radiative
population modelling for hydrogen.

\end{abstract}


\section{Introduction}
\label{sec:intro}

Theoretical modelling of the observed recombination spectra of H~{\sc i} and He~{\sc ii} (and some metals) 
is an important tool for predicting temperatures, densities, abundances (and more) of the local 
emitting/absorbing environment and thereby the testing of large-scale models of said environments. 
These range through  gaseous nebulae (Osterbrock \& Ferland 2006), 
H~{\sc ii} regions (Anderson \etal 2018, Morabito \etal 2014),
active galactic nuclei (Scotville \& Murchikova 2013), the interstellar medium (Oonk \etal 2017)
as well as the early universe (Izotov \etal 2007, 2014).

There has been an upsurge in interest in $l$-changing angular momentum collisions in recent years 
as ever greater precision is being demanded of spectral diagnostics.
The capture-cascade problem is relatively straightforward to model. But heavy-particle collisions
are efficient at changing the $l$-distribution of Rydberg atomic states during the cascade process 
and thus the intensity of lower-lying spectral diagnostic lines.

The seminal paper by Pengelly \& Seaton (1964) used impact parameter theory to describe 
$l$-changing collisions. They provided simple analytic expressions for cross sections and 
rate coefficients for modelling use. All was quiet for half a century.
Then Vrinceanu \& Flannery (2001) analytically solved the time-dependent Schr\"{o}dinger equation
for a colliding heavy particle
creating a weak electric field which lifts the Stark degeneracy in Rydberg atomic states.
Comparison with quantum mechanical (QM) rate coefficients from this method
showed that the simple expression of Pengelly \& Seaton (1964)
was not sufficiently accurate in extreme cases such as low temperatures (Guzm\'{a}n \etal 2016) and/or
for non-degenerate transitions (Guzm\'{a}n \etal 2017).

Evaluation of the analytic quantum mechanical rate coefficients is rather demanding for modelling codes
to carry-out routinely. Improved simple analytic expressions were sought.
Guzm\'an \etal (2017) introduced a modified version of the Pengelly \& Seaton (1964) approach
which improved the description of close encounters (small impact parameters). 
Simple analytic expressions were resultant still which described both dense plasmas 
and non-degenerate transitions separately c.f. Pengelly \& Seaton (1964). This is necessary
because the dipole $l$-changing collision rate coefficients are logarithmically divergent 
unless an environmental cut-off is applied to the contribution from distant encounters 
(large impact parameters). The dense plasma (Debye) cut-off is
independent of the collider energy but cut-offs due to non-degeneracy of a Rydberg transition 
or finite lifetime of the Rydberg state depend on the collider energy alone (Pengelly \& Seaton 1964;
Guzm\'an \etal 2017).

Vrinceanu \etal (2017, 2019) introduced a semi-classical (SC) approximation which gave an
improved description at small impact parameters. The price to pay was in obtaining 
an analytic expression for the rate coefficient. Vrinceanu \etal (2019) provided one for
the case of Debye cut-off. To do so requires that the description of the collision
problem does not depend independently on the impact parameter and collider energy.
This enables a single (combined impact parameter/energy) analytic integration
of the probability to be carried-out so as to deliver a rate coefficient.
The combined dependency is true in principle but the introduction of an energy-dependent cut-off nullifies it.
Energy-dependent cut-offs are important for non-hydrogenic targets and low-lying atomic p-states.
The analytic modified Pengelly \& Seaton rate coefficients of Guzm\'{a}n \etal (2017) are required here.
It is a concern then that Vrinceanu \etal (2019) appeared to obtain very poor results from the
modified Pengelly \& Seaton rate coefficients of Guzm\'{a}n \etal (2017) for proton collisions with
hydrogen.

In Section \ref{sec:PSM} we review, clarify and improve-upon the modified Pengelly \& Seaton (PSM)
method introduced by Guzm\'an \etal (2017).
We show good accord between correctly interpreted PSM, semi-classical and quantum mechanical results
in Section \ref{sec:results}.
We summarize our findings in Section \ref{sec:concludes}.
In Appendix \ref{sec:appendix} to this paper we detail the efficient 
numerical algorithms that we use to evaluate the quantum mechanical probabilities first formulated by 
Vrinceanu \& Flannery (2001).

\section{Methodology}
\label{sec:PSM}

Impact parameter theory (Alder \etal 1956) can be used to write the cross section $\sigma_{ji}$
for an atomic transition $i\rightarrow j$ as
\be
\sigma_{ji} = 2\pi \int^\infty_0 P_{ji}(R) R \rmd R
\label{Sigma_ij}
\ee
in terms of the transition probability $P_{ji}(R)$ and impact parameter $R$.

The Bethe approximation can be used to write the probability for dipole transitions 
($l\rightarrow l'=l\pm 1$) as
\be
P_{ji}(R)=\frac{a_0^2\mu I_\rmH}{2\omega_l E}\frac{D_{ji}}{R^2}
\label{P_ij}
\ee
where $E$ is the energy of the collider in units $I_\rmH$, $\mu$ is the dimensionless reduced mass of the 
target--collider system, $\omega_l=2l+1$ and $a_0$ is the Bohr radius.

The dipole factor $D_{ji}$ for $l$-changing collisions 
(which is closely related to the atomic line strength) is given by
\be
D_{ji}=\frac{Z^2}{z^2} 6n^2 l_>(n^2-l^2_>)
\label{D_ij}
\ee
where $Z$ is the charge of the collider, $z$ is the charge of the target as seen by the Rydberg electron
$nl$ and $l_>=\max(l,l')$.

Energy-degenerate dipole transitions give rise to a logarithmic divergence in the cross section due 
to the contribution from distant encounters. The standard approach (Pengelly \& Seaton, 1964) 
is to introduce a large impact parameter cut-off at
$R_\rmc$. The cut-off due to a finite density plasma neutralizing a Debye sphere is independent of the
energy of the colliding particle. Finite lifetimes of the excited target-states and  non-degenerate
target energies for the transition clearly lead to a (collider) energy-dependent cut-off.
The quantum mechanical (Vrinceanu \& Flannery 2001) and semi-classical approximations of 
Vrinceanu \etal (2017, 2019) require such a cut-off as well.

Use of equation (\ref{P_ij}) also gives rise to a divergent probability as $R\rightarrow 0$.
Pengelly \& Seaton (1964) introduced a critical small impact parameter $R_1$ below which the
probability was bounded: $P_{ji}(R<R_1)=P_1$ say. This completes the definition
of the final-state resolved Pengelly \& Seaton (1964) approximation. We denote it PS64.

It is well known that the PS64 approximation gives poor results and eventually breaks down for
problems dominated by the contribution from small impact parameters such as low temperatures
and/or high densities and/or severely non-degenerate transition energies.

%%%%%%%%%%%%%%%%%%%%%%%%%%%%%%%%%%%%%%%%%%%%%%%%%%%%%%%%%%%%%%%%%%%%%%%%%%%%%%%%%%%%%%%%%%%%%%%%%%%%%%%%%
\begin{table*} \footnotesize
\begin{center}
\caption{\label{t:rates19} 
%\footnotesize
Comparison of He--p rate coefficients $q_{nl\rightarrow nl'}$  ($\text{cm}^{3}\text{s}^{-1}$) from
the different theoretical PS, SC and QM methods for $n=30$ and low- and high-$l$ for
different temperatures $(T_\rmH)$ at a hydrogen density $N_\rmH$ of $100$~cm$^{-3}$.  
QM-VOS12 denotes our use of the QM formula given in Vrinceanu \etal (2012);
PS64 is the `standard' Pengelly \& Seaton (1964);
PSM17 uses the original $P\propto R/R_c$ of Guzm\'an \etal (2017) when $R\le R_\rmc\le R_1$ while 
PSM20 uses the present $P\propto R/R_1$ instead;
SC-VOS17 denotes the semi-classical method of Vrinceanu \etal (2017) and the results from which 
were not available to Guzm\'an \etal (2017).
 }
%\vspace{5mm}
\begin{tabular}{c c c c c  |}
\hline
%\cline{3-5}
 & & \multicolumn{3}{|c|}{$N_\text{H}=100\text{ cm}^{-3}$}\\
%\cline{3-5}
& & \multicolumn{1}{|c|}{$T_\rmH=10^2\text{K}$}&\multicolumn{1}{|c|}{$T_\rmH=10^4\text{K}$}
&\multicolumn{1}{|c|}{$T_\rmH=10^6\text{K}$}\\
\hline
\multicolumn{1}{|c}{$l=4 \to l^\prime=3$} 
& \multicolumn{1}{|c|}{QM-VOS12}                        & $1.66[-3]\dag$ & $5.61[+0]$ & $3.51[+0]$ \\
\multicolumn{1}{|c}{} & \multicolumn{1}{|c|}{PS64}      &  ---       & $4.18[+0]$ & $3.65[+0]$ \\
\multicolumn{1}{|c}{} & \multicolumn{1}{|c|}{PSM17}     & $2.00[-2]$ & $5.77[+0]$ & $3.57[+0]$ \\
\multicolumn{1}{|c}{} & \multicolumn{1}{|c|}{PSM20}     & $1.24[-3]$ & $5.70[+0]$ & $3.57[+0]$ \\
\multicolumn{1}{|c}{} & \multicolumn{1}{|c|}{SC-VOS17}  & $1.91[-3]$ & $6.25[+0]$ & $3.94[+0]$ \\
\hline
\multicolumn{1}{|c}{$l=29\to l^\prime=28$} 
& \multicolumn{1}{|c|}{QM-VOS12}                        & $3.80[+1]$ & $6.18[+0]$ & $8.55[-1]$ \\
\multicolumn{1}{|c}{} & \multicolumn{1}{|c|}{PS64}      & $4.06[+1]$ & $6.44[+0]$ & $8.81[-1]$ \\
\multicolumn{1}{|c}{} & \multicolumn{1}{|c|}{PSM17}     & $3.80[+1]$ & $6.18[+0]$ & $8.54[-1]$ \\
\multicolumn{1}{|c}{} & \multicolumn{1}{|c|}{PSM20}     & $3.80[+1]$ & $6.18[+0]$ & $8.54[-1]$ \\
\multicolumn{1}{|c}{} & \multicolumn{1}{|c|}{SC-VOS17}  & $3.84[+1]$ & $6.26[+0]$ & $8.67[-1]$  \\
\hline
\dag $1.66[-3]$ denotes $1.66\times 10^{-3}$.
\end{tabular}
\end{center}
\end{table*}
%%%%%%%%%%%%%%%%%%%%%%%%%%%%%%%%%%%%%%%%%%%%%%%%%%%%%%%%%%%%%%%%%%%%%%%%%%%%%%%%%%%%%%%%%%%%%%%%%%%%%%%%%

Guzm\'an \etal (2017) introduced a modification of PS64 to overcome this limitation. It is based
upon the behaviour of the quantum mechanical probability (Vrinceanu \& Flannery, 2001; Vrinceanu \etal, 2012)
at small impact parameters.
They chose
\be
P_{ji}(R<R_1)=P_1 \frac{R}{R_1}\,.
\label{PPSM}
\ee
Combining (\ref{PPSM}) with equation (\ref{P_ij}) leads to the matching condition which defines $R_1$:
\be
P_1R^2_1=\frac{a_0^2\mu I_\rmH}{2\omega_l E}D_{ji}\,.
\label{Ronem}
\ee
The cross section is then given by
\be
\sigma_{ji}(E)=
\pi P_1 R^2_1 \left[\frac{2}{3}+2\ln\left(\frac{R_\rmc}{R_1}\right)\right]\quad\quad  \mbox{when $R_\rmc\ge R_1$}
\label{QPSM1a}
\ee
and by
\be
\sigma_{ji}(E)=
\pi P_1 R^2_1 \left(\frac{R_\rmc}{R_1}\right)^{3}\,{\frac{2}{3}} \quad\quad \mbox{when  $R_\rmc<R_1$}\,.
\label{QPSM1b}
\ee
%\be
%\sigma_{ji}=\left\{\begin{array}{ll}
%\pi P_1 R^2_1 \left[\frac{2}{3}+2\ln\left(\frac{R_\rmc}{R_1}\right)\right] & \mbox{when $R_1\le R_\rmc$}\\
%\pi P_1  \frac{R^3_\rmc}{R_1}{\frac{2}{3}} & \mbox{when $R_1>R_\rmc$}\,.\\
%\end{array}\right.
%\label{QPSM1}
%\ee
The cross sections for $R_\rmc<R_1$ (equation \ref{QPSM1b}) correspond with
the scattering energies $E<E_\min$:
\be
E_\min=\frac{a_0^2\mu I_\rmH}{2P_1\omega_l R_\rmc^2}D_{ji}
\label{Emin}
\ee
which is defined by setting $R_1=R_\rmc$ in equation (\ref{Ronem}). Cross sections at these energies
are neglected by PS64.
We denote this approximation PSM.

The corresponding rate coefficient $q_{ji}$ at an electron temperature $T_\rme$
is obtained by convoluting the cross section with a Maxwellian distribution over {\it all}
collider energies. It takes on two forms.

(1) If the cut-off $R_\rmc$ is independent of the collider energy (e.g. Debye) then
\be
q_{ji}= \frac{a^3_0}{\tau_0}\left(\frac{\pi \mu I_\rmH}{\kb T_\rme}\right)^\half \frac{D_{ji}}{\omega_l}
\left[
\frac{\sqrt{\pi}}{2}U_\rmm^{-\3half} \mbox{erf}(U_\rmm^{\half})-\rme^{-U_\rmm}/U_\rmm+E_1(U_\rmm)
\right]
\label{QPSM3+EMIN}
\ee
where erf() denotes the error function, $E_1()$ the first exponential integral, $U_\rmm=E_\min/\kb T_\rme$,
$\kb$ the Boltzmann constant and $\tau_0$ the Bohr time.

Guzm\'an \etal (2017) did not give this complete Debye form of the PSM rate coefficient since they
were studying helium and so required the use of an energy dependent cut-off.

(2) The energy-dependent lifetime/splitting cut-off $R_\rmc(E)\propto \sqrt{E}$ will always
be larger than the Debye one at sufficiently large collider energies.
Guzm\'an \etal (2017) discuss how to split the convolution into two energy ranges $[0,E_\rmc]$
and $[E_\rmc,\infty)$ where the energy $E_\rmc$ is defined by $R_\rmc(E)=R_\rmc(\mbox{Debye})$.
Thus
\be
R^2_\rmc(E_\rmc)=\frac{E_\rmc t^2}{I_\rmH \mu}=\frac{\kb T_\rme}{8\pi a_0 I_\rmH N_\rme}=R^2_\rmc(\mbox{Debye})
\ee
and so
\be
E_\rmc=\frac{\mu\kb T_\rme}{8\pi a_0 t^2 N_\rme}\,.
\ee
Here $N_\rme$ is the electron density (which defines the Debye sphere) and $t$ is written in
terms of the lifetime of the upper state ($\tau_{nl}$) or in terms of the energy splitting ($\Delta E_{ji}$)
%
% gary gjf comment out
for the transition viz. $t=0.72\tau_{nl}$ or $t=1.12\hbar/\Delta E$ --- see
Pengelly \& Seaton (1964), Guzm\'an \etal (2017).

The rate coefficient in this case is
\beq
\label{QPSM4+EMIN}
q_{ji}
&\hspace{-3mm}=&\hspace{-3mm}\frac{a^3_0}{\tau_0}\left(\frac{\pi \mu I_\rmH}{\kb T_\rme}\right)^\half 
\frac{D_{ji}}{\omega_l}\\
&\hspace{-4mm}\times&
\hspace{-4mm}\left[4\left\{1-\rme^{-\Ubar_\rmm}\left(1+\Ubar_\rmm
%\right.\right.\right.\nonumber\\
%&+& 
% \left. \left. \left.
+\tfrac{1}{2}\Ubar_\rmm^2\right)\right\}\Ubar_\rmm^{-3}
+2E_1(\Ubar_\rmm)-E_1(U_\rmc)\right]\nonumber
\eeq
where $\Ubar_\rmm^2=U_\rmm U_\rmc$ and $U_\rmc=E_\rmc/\kb T_\rme$. 
$E_1(U_\rmc\rightarrow \infty) \rightarrow 0$ applies the energy dependent cut-off at 
all energies. Note that this formula (\ref{QPSM4+EMIN}) assumes that $U_\rmc\ge \Ubar_\rmm$.
A more tedious expression results otherwise. We have yet to encounter its need.

%%%%%%%%%%%%%%%%%%%%%%%%%%%%%%%%%%%
\begin{figure}
\vspace{-80mm}
\hspace{-21mm}\includegraphics[angle=0, scale=.75]{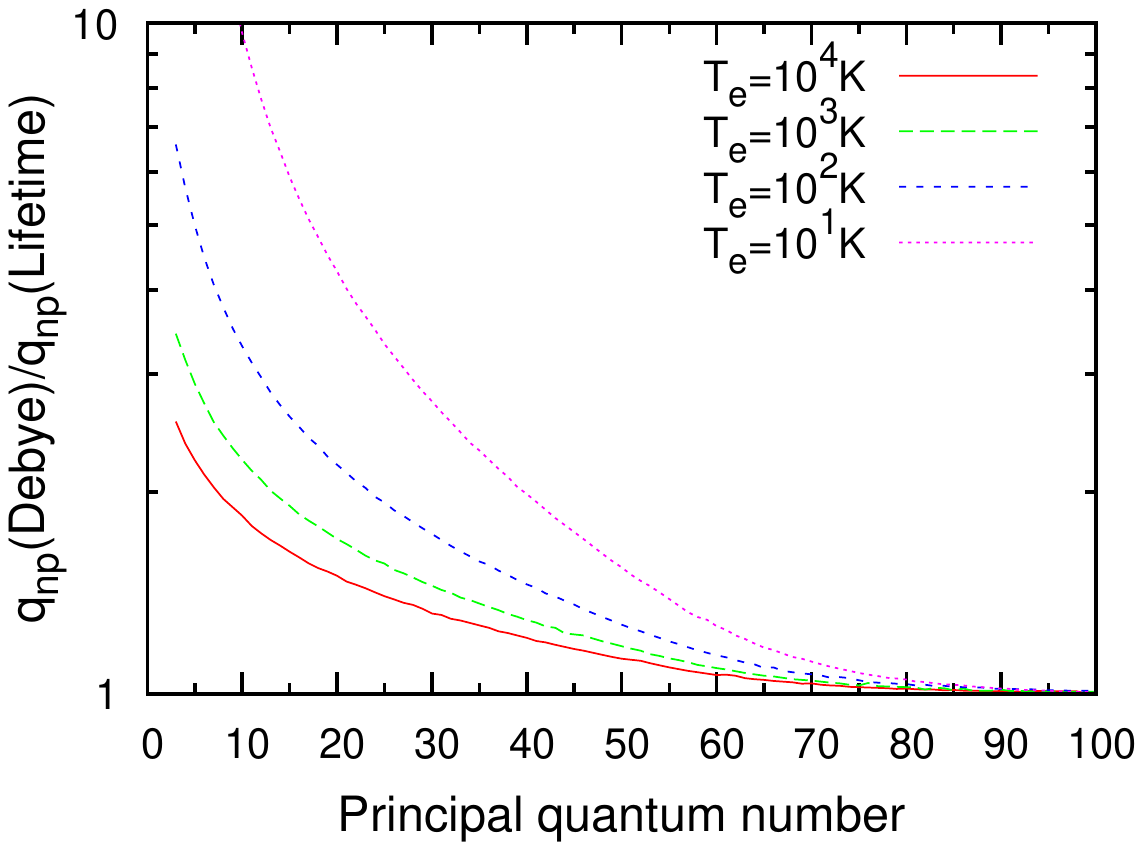}
\vspace{-14mm}
\caption{
Ratio of PSM Debye-to-lifetime cut-off H--p rate coefficients $q_{n\rmp}$  at 
$T_\rme=10^1, 10^2, 10^3, 10^4$~K and $N_\rme=100$ cm$^{-3}$.}
\label{fig1}
\end{figure}
%%%%%%%%%%%%%%%%%%%%%%%%%%%%%%%%%%%

Figure \ref{fig1} shows the importance of applying a lifetime cut-off rather than a Debye cut-off for 
low-lying $n$p states in H--p collisions.

Neither equation (\ref{QPSM4+EMIN}) nor equation (\ref{QPSM3+EMIN}) correspond
quite with those discussed by Guzm\'an \etal (2017). The reason for this is that
Guzm\'an \etal (2017) switched to using
\be
P_{ji}(R)=P_1 \frac{R}{R_c}
\label{PPSMC}
\ee
for $R\le R_c \le R_1$.
This leads to 
\be
\sigma_{ji}(E\rightarrow 0) \sim E^0\quad  \mbox{or} \quad E^1
\label{QPSMC}
\ee
for $R_c$(Debye)  or $R_c(E)$ respectively.

Use of equation (\ref{PPSM}) still for $R\le R_c \le R_1$ leads to (see equations (\ref{Ronem}) \& (\ref{QPSM1b}))
\be
\sigma_{ji}(E\rightarrow 0) \sim E^{1/2}\quad  \mbox{or} \quad E^2
\label{QPSMC1}
\ee
for $R_c$(Debye)  or $R_c(E)$ respectively.

Study of the quantum mechanical cross sections at low energies does not yield an
obvious verdict as to which to use. The asymptotic form does not appear to be reached until such low
energies as to be irrelevant for temperatures $>1$~K. The behaviour of the cross section at higher
non-asymptotic energies likely dominates the practical application. There is some evidence from He--p 
collisions that the use of equation (\ref{PPSM}) is preferable still.
In the present Table \ref{t:rates19} we re-visit the He--p problem whose results were shown in Table 1 
of Guzm\'an \etal (2017).
The PSM results shown by Guzm\'an \etal (2017) broke down in the extreme case of a highly non-degenerate 
transition at low temperature.  (We denote them PSM17.) The present results (which we denote PSM20)
are of comparable accuracy to those we have obtained using the semi-classical approximation
of Vrinceanu \etal (2017) and which were not available to Guzm\'an \etal (2017).

We return now to $P_1$ which bounds the probability for close encounters.
We define generally
\be
P_1=\frac{1}{2}B_{ji}
\label{P1B}
\ee
where the branching ratio $B_{ji}$ is given 
\be
B_{ji}=\frac{D_{ji}}{\omega_l D_{nl}}
\ee
and $D_{nl}$ is the unresolved dipole quantity used by PS64
\be
D_{nl}=\sum_{l'=l\pm 1} \frac{1}{\omega_l}D_{ji}=\frac{Z^2}{z^2}6n^2(n^2-l^2-l-1) \,.
\label{DPS}
\ee
Guzm\'an \etal (2017) compared their PSM probabilities with the results that they obtained from the quantum mechanical
approach of Vrinceanu \& Flannery (2001) and detailed by Vrinceanu \etal (2012). Guzm\'an \etal (2017) chose
\be
P_1=\frac{1}{4}\,.
\label{P14}
\ee
This is in contrast to  Summers (1977) and Hummer \& Storey (1987) who retained equation (\ref{P1B})
to define their constant bound. The improvement on using equation (\ref{P14}) is marginal.
We note that summing-over both final states leads in both cases to
\be
P_1=\frac{1}{2}
\ee
which is in agreement with Pengelly \& Seaton (1964).

Comparison of the total collisional rate ($N_\rme q_{nl}$) out of a state ($nl$) with the total radiative
rate out ($A_{nl}$) is of interest in population modelling: $N_\rme q_{nl}\tau_{nl}=1$ defines the
critical density above which $nl\rightarrow nl'$ collisions are faster than radiative ones ($\tau_{nl}=1/A_{nl}$).

We have formulated $l$-changing collisions in a final-state resolved picture.
One can simply sum over the final-state resolved rate coefficients
\be
q_{nl}=\sum_{l'=l\pm 1}q_{nl\rightarrow nl'}
\ee
to obtain a total unresolved rate coefficient.
This is the only procedure available in the quantum mechanical case.

 The problem was formulated historically in an unresolved picture ---
recall the original Pengelly \& Seaton formula. Vrinceanu \etal (2019)
consider an unresolved picture.
By unresolved picture we mean that the matching point (e.g. $R_1$) in Pengelly \& Seaton ($\pm$ modified) and 
the semi-classical approach of Vrinceanu \etal (2019) is defined in terms of the {\it total} probability out of $nl$.
The modified Pengelly \& Seaton $P_{nl\rightarrow nl-1}$ and $P_{nl\rightarrow nl+1}$
have {\it different} matching points in the resolved picture since we take $B_{ij}=1/2$.

At this point it is worth recalling that the modified Pengelly \& Seaton method
was optimized for $nl\rightarrow nl-1$ transitions and data for $nl\rightarrow nl+1$ transitions
should be determined from $nl\leftarrow nl+1$ via reciprocity e.g.
\be
q_{nl}=q_{nl\rightarrow nl-1}+\frac{(2l+3)}{(2l+1)}q_{nl+1\rightarrow nl}\,.
\ee
This (application of reciprocity) is the normal procedure for evaluating
all rates and rate coefficients when carrying-out
population modelling so as to ensure one attains the LTE limit at high density.

The unresolved and resolved modified Pengelly \& Seaton approaches should give similar 
results and increasingly so as the contribution from small impact parameters lessens.
The contribution from small impact parameters becomes important
at low temperatures and/or high densities.
The original Pengelly \& Seaton approach starts to fail here. The flexibility
of using different resolved matching points may offer some improvement over the unresolved approach.

It is simple to deduce the modified Pengelly \& Seaton formulae for the {\it unresolved} picture from
the ones already given for the {\it resolved} picture:

1/ Replace $B_{ij}$ by unity: {\bf thus} \boldmath${P_1=1/2}$ {\bf here} e.g. in Equ. (\ref{Emin}).

2/ Replace $D_{ji}/\omega_l$ by $D_{nl}$.
\newline
We note that simply summing over the final-states in the resolved picture will yield (somewhat)
different results to those obtained from using the explicit unresolved formulae of
the modified Pengelly \& Seaton approach.
Both approaches require the evaluation of $\sim n$ expressions
of similar complexity and so are similar in terms of computational effort.

%%%%%%%%%%%%%%%%%%%%%%%%%%%%%%%%%%%
\begin{figure}
\vspace{-80mm}
\hspace{-21mm}\includegraphics[angle=0, scale=.75]{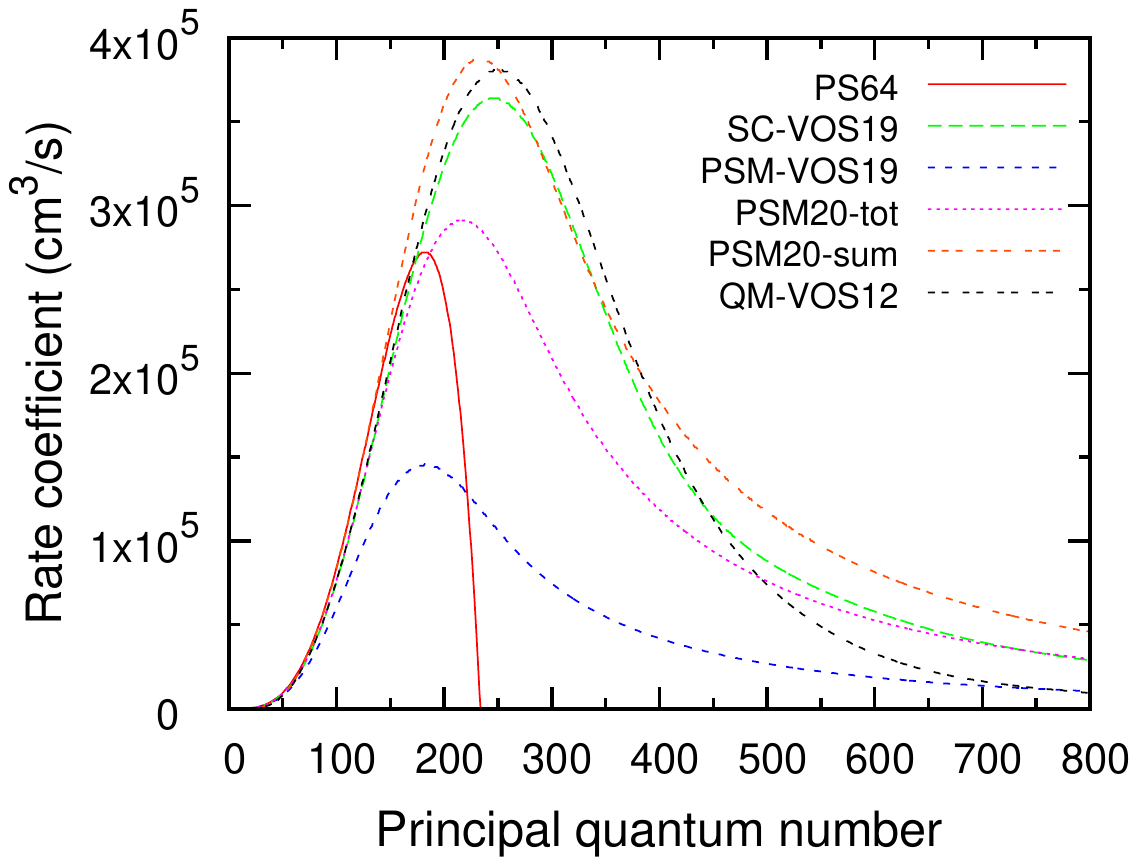}
\vspace{-14mm}
\caption{
Total H--p rate coefficients $q_{nl}$ for $n, l=1$ at $T_\rme=10$~K and $N_\rme=100$ cm$^{-3}$.
PS64 denotes the `standard' Pengelly \& Seaton (1964) method;
SC-VOS19 denotes the semi-classical method of Vrinceanu \etal (2019);
PSM-VOS19 denotes the modified Pengelly \& Seaton method ($P_1=1/4$) of Vrinceanu \etal (2019);
PSM20 denotes the present modified Pengelly \& Seaton method ($P_1=1/2$);
`tot' denotes unresolved; 'sum' denotes resolved-sum (see text);
QM-VOS12 denotes our use of the QM expressions given by Vrinceanu \etal (2012).}
\label{fig2}
\end{figure}
%%%%%%%%%%%%%%%%%%%%%%%%%%%%%%%%%%%

%%%%%%%%%%%%%%%%%%%%%%%%%%%%%%%%%%%
\begin{figure}
\vspace{-80mm}
\hspace{-21mm}\includegraphics[angle=0, scale=.75]{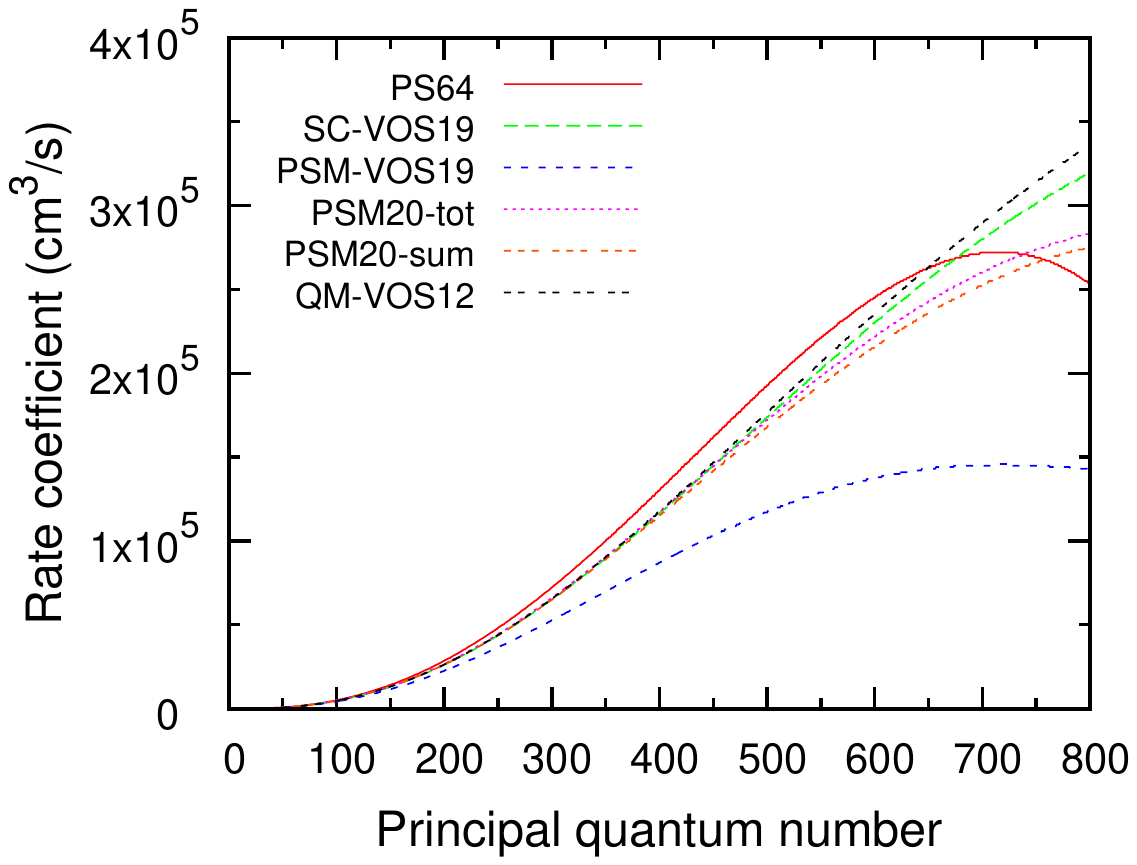}
\vspace{-14mm}
\caption{
Total H--p rate coefficients $q_{nl}$ for $n, l=n-2$ at $T_\rme=10$~K and $N_\rme=100$ cm$^{-3}$.
PS64 denotes the `standard' Pengelly \& Seaton (1964) method;
SC-VOS19 denotes the semi-classical method of Vrinceanu \etal (2019);
PSM-VOS19 denotes the modified Pengelly \& Seaton method ($P_1=1/4$) of Vrinceanu \etal (2019);
PSM20 denotes the present modified Pengelly \& Seaton method ($P_1=1/2$);
`tot' denotes unresolved; 'sum' denotes resolved-sum (see text);
QM-VOS12 denotes our use of the QM expressions given by Vrinceanu \etal (2012).}
\label{fig3}
\end{figure}
%%%%%%%%%%%%%%%%%%%%%%%%%%%%%%%%%%%

\section{Results}
\label{sec:results}

All results shown in this section are calculated using the appropriate Debye cut-off unless
stated otherwise.

Figure \ref{fig2} compares total $l$-changing rate coefficients out of $n\rmp$ states in hydrogen at
an electron temperature of 10~K and density 100~cm$^{-3}$. A similar comparison was shown
by Vrinceanu \etal (2019). They highlighted the poor agreement of the modified Pengelly \& Seaton
results (which we denote PSM-VOS19) with all other methods. 
This is due to the incorrect use by Vrinceanu \etal (2019) of $P_1=1/4$ for an unresolved transition.
The correct results 
obtained using $P_1=1/2$ (which we denote PSM20-tot) are in much better accord.
We note that we have not attempted to re-optimize the PSM $R_1$ matching point for this problem.
Vrinceanu \etal (2019) re-optimized their matching point compared to Vrinceanu \etal (2017).
Rather better agreement is found for PSM for $n$-values where the rate coefficient is largest
if we sum-over the resolved rate coefficients (which we denote PSM20-sum).
Figure \ref{fig3} makes a similar comparison as Figure \ref{fig2} but now for $n,l=n-2$.
The results of all methods are in close accord except for the starkly different PSM-VOS19 ones.

Vrinceanu \etal (2019) present QM rate coefficients calculated at 15 $n$-values in their
Figures 2(a) and 2(b). These correspond to our Figures \ref{fig2} and \ref{fig3}.
Vrinceanu \etal (2019) state that those QM rate coefficients for $n\rmp$ took
several hours of CPU time while those for $n,l=n-2$ took 2 days.
We calculated our corresponding QM results at 800 $n$-values in less than 10s and 30s respectively.
We detail in the Appendix the fast and efficient numerical algorithms that we have implemented.
and which only require standard 64-bit floating point arithmetic.
The algorithms used by Vrinceanu \etal (2019) required 400 digits of precision.
The 5 orders of magnitude speed-up that we obtain with our algorithms means that their efficient implementation
within modelling codes should make calculations using the QM method much more routine.

%%%%%%%%%%%%%%%%%%%%%%%%%%%%%%%%%%%
\begin{figure}
\vspace{-75mm}
\hspace{-14mm}\includegraphics[angle=0.75, scale=.7]{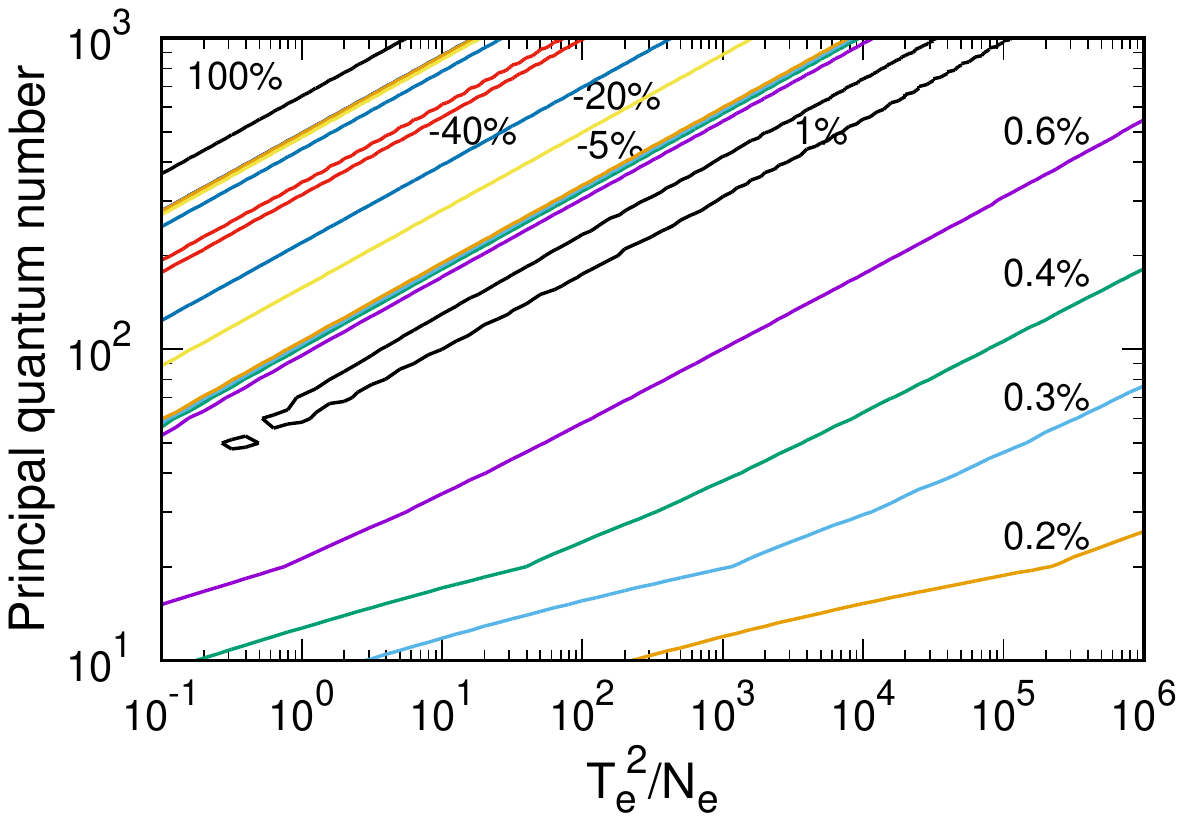}
\vspace{-17mm}
\caption{
Percentage difference between PSM20-tot and QM-VOS12 total H--p rate coefficients $q_{nl}$ for $n, l=1$
as a function of $T_\rme^2/N_\rme$.
PSM20-tot denotes the present unresolved modified Pengelly \& Seaton method;
QM-VOS12 denotes our use of the QM expressions given by Vrinceanu \etal (2012).}
\label{fig4}
\end{figure}
%%%%%%%%%%%%%%%%%%%%%%%%%%%%%%%%%%%

Figure \ref{fig4} shows the percentage difference between our PSM20-tot results and the QM results
(which we denote QM-VOS12) that we have computed using the expressions given by Vrinceanu \etal (2012).
The comparison is again made for $n\rmp$ states in hydrogen at an electron temperature of 10~K and density 100~cm$^{-3}$.
We see that PSM20-tot is accurate to 1\% or better over a wide range of the Debye temperature--density
parameter space. This is in contrast to the few percent difference illustrated by Vrinceanu \etal (2019)
for the original Pengelly \& Seaton results (PS64). Vrinceanu \etal (2019) showed that their semi-classical results
were also accurate to better than 1\% over a similar range of parameter space --- typically a factor 2
more accurate for a given temperature--density. All simple methods breakdown rapidly at a critical
and similar temperature--density diagonal ($T_\rme^2/(N_\rme n^4)$) as seen in Figure \ref{fig4}.
The PSM results are well-behaved for all $T_\rme^2/(N_\rme n^4)$. They dip down and underestimate by up to 40\% but 
ultimately end-up as a large overestimate compared to the QM rate coefficients.
But the QM rate coefficients themselves are very large by then. Both sets of rates have already established a 
statistical $l$-population. Their magnitude is no longer relevant. Guzm\'an \etal (2016, 2017) provide illustrative
figures for the H {\sc i} and He {\sc i} recombination spectra. All methods agree at low and high densities 
(excluding the original PS64).

We note that the results and timings for Figure \ref{fig2} correspond to single vertical line in Figure \ref{fig4}.
We have created and examined contour plots similar to those of Figure \ref{fig4} but for $l=n/2$ and $l=n-2$.
They all show a similar pattern. The results shown in Figure \ref{fig4} are thus representative of the $l$-space as well.

%%%%%%%%%%%%%%%%%%%%%%%%%%%%%%%%%%%
\begin{figure}
\vspace{-80mm}
\hspace{-21mm}\includegraphics[angle=0, scale=0.75]{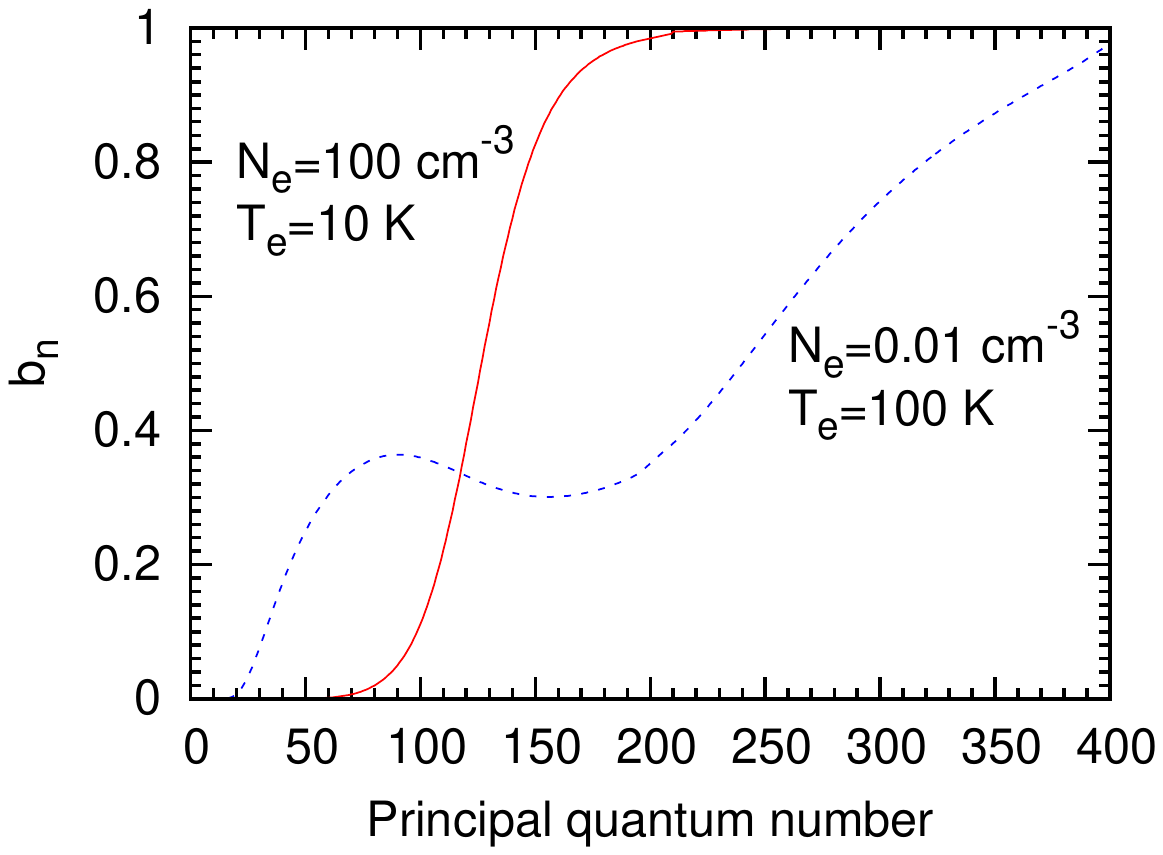}

\vspace{-90mm}
\hspace{-21mm}\includegraphics[angle=0, scale=0.75]{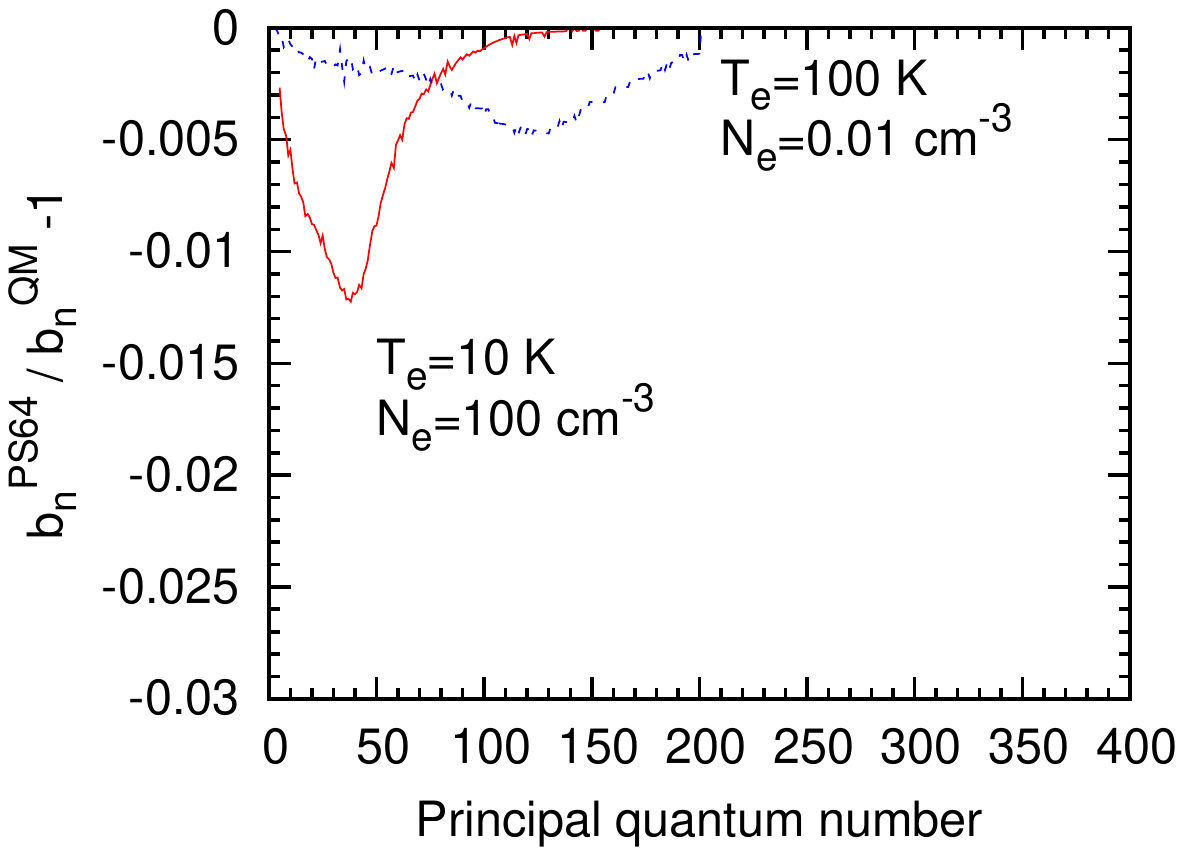}
\vspace{-14mm}
\caption{
Hydrogen population modelling at $T_\rme=100$~K \& $N_\rme=0.1$~cm$^{-3}$ and
$T_\rme=10$~K \& $N_\rme=100$~cm$^{-3}$.
Upper: departure coefficients $b_n$. The results for the present unresolved modified 
Pengelly \& Seaton method (PSM20-tot), the original Pengelly \& Seaton (1964) method (PS64) and
our use of the QM expressions given by Vrinceanu \etal (2012) (QM-VOS12) are indistinguishable.
Lower: fractional differences between PS64 and QM-VOS12 departure coefficients ($b^{\rm PS64}_n/b^{\rm QM}_n-1$). 
The differences between PSM20-tot and QM-VOS12 are not distinguishable from zero in this figure.}
\label{fig5}
%\vspace{-10mm}
\end{figure}
%%%%%%%%%%%%%%%%%%%%%%%%%%%%%%%%%%%

We have carried-out population modelling with the spectral simulation code {\sc cloudy} (Ferland \etal 2017).
We used revision r13930M on the PSM20 development branch of {\sc cloudy} in
which we have implemented the new equations given in Section \ref{sec:PSM}.
We again used the hydrogen-only-cloud model described by  Guzm\'an \etal (2016). 
Figure \ref{fig5} (upper) shows the thermal departure coefficients ($b_n$) at
$T_\rme=100$~K \& $N_\rme=0.1$~cm$^{-3}$ and $T_\rme=10$~K \& $N_\rme=100$~cm$^{-3}$. 
The $b_n$ calculated using the  PSM20-tot, PS64 and QM-VOS12 methods are indistinguishable in this figure.
Figure \ref{fig5} (lower) shows the corresponding fractional differences between PS64 and QM-VOS12 ($b^{\rm PS64}_n/b^{\rm QM}_n-1$).
The largest difference is $\sim 0.5\%$ and $\sim 1.2\%$ respectively for the two $(T_\rme,N_\rme)$ cases.
Vrinceanu \etal (2019) carried-out a similar comparison of their semi-classical results with the results of the original
Pengelly \& Seaton (1964) formula (PS64). The maximal differences were $\sim 0.8\%$ and $\sim 3.2\%$ for the same two cases.
It should be noted that the two hydrogen models differ in their large-scale description (see Guzm\'an \etal (2016)
and Vrinceanu \etal (2019) for details). 
The PSM20-tot fractional differences with QM-VOS12 are very small ($\sim 10^{-6}$)
as are those for the original PSM17 approximation of Guzm\'an \etal (2017).

\section{Conclusions}
\label{sec:concludes}

We have reviewed, clarified and improved-upon the modified Pengelly \& Seaton PSM 
method introduced by Guzm\'an \etal (2017) which describes atomic $l$-changing collisions.

\begin{itemize}

\item 
We have shown that an alternative treatment of small impact parameters leads to much
improved results from PSM in extreme cases such as highly non-degenerate transitions in He.
We have provided complete expressions for rate coefficients for both an energy independent
(Debye) cut-off at large impact parameters (see equation (\ref{QPSM3+EMIN})) as well as for (collider) 
energy-dependent cut-offs due to non-degenerate transitions and/or finite atomic lifetimes
(see equation (\ref{QPSM4+EMIN})).
The latter is not available for the semi-classical approach of Vrinceanu \etal (2017, 2019).

\item
We have pointed-out the mis-interpretation made by Vrinceanu \etal (2019) when they
adapted the final-state resolved PSM approach of Guzm\'an \etal (2017) to the unresolved case.
Correct interpretation leads to good accord between PSM  results and those we have obtained using their
semi-classical  (Vrinceanu \etal 2019) and quantum mechanical (Vrinceanu \& Flannery 2001) methods.

\item
We have described the numerical algorithms that we use to evaluate the quantum mechanical probabilities
(see Appendix \ref{sec:appendix}).
They are many orders of magnitude faster than those described by Vrinceanu \etal (2019) and
they only require the use of standard 64-bit floating point arithmetic.
Their efficient implementation within modelling codes should make such calculations much more routine.

\end{itemize}

\section{Data Availability}
Results shown in the Figures will be shared on reasonable request to the corresponding author.

The Fortran codes which implement the algorithms described in Appendix A are available from the
UK APAP Network website: apap-network.org. This includes a library of Wigner 3n-j programs
(at apap-network.org/3n-j) and a test-driver/wrapper-plus-subprogram to calculate the QM and PSM20
Maxwellian rate coefficients (at apap-network.org/lchng). The programs are interactive and should be
self-explanatory.

\section{Acknowledgements}
NRB acknowledges support from STFC (UK) through the University of Strathclyde APAP Network grant ST/R000743/1. 
FG and GJF acknowledge support from the National Science Foundation 
(grant number 1816537) and NASA ATP program (grant number 17-ATP17-0141). 
MC acknowledges support by NSF (1910687), NASA (19-ATP19-0188), and 
STScI (HST-AR-14556.001-A).

\section{References}
Alder K., Bohr A., Huus T., Mottelson B., Winther A., 1956, Rev.Mod.Phys., 28, 432 \\
Abramowitz M., Stegun I. A., 1972, {\it Handbook of Mathematical Functions} Dover, NY \\
Anderson L. D., Armentrout W. P., Luisi M., Bania T. M., Balser D. S., Wenger T. V., 2018, ApJS, 234, 33 \\
Edmonds A. R., 1957 {\it Angular Momentum in Quantum Mechanics} Princeton, NJ \\
Ferland G. J., Chatzikos M., Guzm\'{a}n F., Lykins M. L., van Hoof P. A. M., Williams R. J. R., Abel N. P., 
Badnell N. R., Keenan F. P., Porter R. L., Stancil P. C., 2017 Rev.Mex.Astron.Astrofis., 53, 385 \\
Guzm\'{a}n F., Badnell N. R., Williams R. J. R., van Hoof P. A. M., Chatzikos M., Ferland G. J., 2016 MNRAS, 459, 3498 \\
Guzm\'{a}n F., Badnell N. R., Williams R. J. R., van Hoof P. A. M., Chatzikos M., Ferland G. J., 2017 MNRAS, 464, 312 \\
Guzm\'{a}n F., Chatzikos M., van Hoof P. A. M., Blaser D. S., Dehghanian M., Badnell N. R., Ferland G. J., 2019 MNRAS, 486, 1003 \\
Izotov Y. I., Thuan T. X., Stasifi\'{n}ska G., 2007, ApJ, 662, 15 \\
Izotov Y. I., Thuan T. X., Guseva N. G., 2014, MNRAS, 445, 778 \\
Luscombe J. H., Luban M., 1998, Phys.Rev.E, 57, 7274 \\
Morabito L. K., Oonk J. B. R.,  Salgado F., Toribio M. C., R\"{o}ttergering H. J. A., Tielens A. G. G. M., Beck R., Adebahr B., Best P.,
Beswick R., Bonafede A., Brunetti G.,  Br\"{u}ggen M., Chy\.{z}y K. T., Conway J. E., van Driel W., Gregson J., Haverkorn M., Heald G., 
Horellou C., Horneffer A., Iacobelli M.,  Jarvis M. J., Marti-Vidal I., Miley G., Mulcahy D. D., Orr\'{u} E., Pizzo R., Scaife A. M. M., 
Varenius E.,van Weeren R. J., White G. J., Wise M. W., 2014, ApJ, 795, L33 \\
Oonk J. B. R., van Weeren R. J., Salas P., Salgado F., Morabito L. K., Toribio M. C., Tielens A. G. G. M., R\"{o}ttergering H. J. A., 2017, MNRAS, 465, 1066 \\
Osterbrock D.E., Ferland G.J.,  2006, Astrophysics of gaseous nebulae and active galactic nuclei, 2nd ed. University Science Books, Sausalito, CA \\
Pengelly R. M., Seaton M. J., 1964, MNRAS, 127, 165 \\
Racah G., 1942, Phys.Rev., 62, 438 \\
Schulten K., Gordon R. G., 1975a, J.Math.Phys., 16, 1961 \\
Schulten K., Gordon R. G., 1975b, J.Math.Phys., 16, 1971 \\
Schulten K., Gordon R. G., 1976, Comput.Phys.Commun., 11, 269 \\
Scoville N., Murchikova L., 2013, ApJ, 779, 75 \\
Summers H. P., 1977, MNRAS, 178, 101 \\
Vrinceanu D.,  Flannery M. R., 2001, Phys.Rev.A 63, 032701; J.Phys.B, 34, L1 \\
Vrinceanu D., Onofrio R., Sadeghpour H. R., 2012, ApJ, 747, 56 \\
Vrinceanu D., Onofrio R., Sadeghpour H. R., 2017, MNRAS, 471, 3051 \\
Vrinceanu D., Onofrio R., Oonk J. B. R., Salas P., Sadeghpour H. R., 2019, ApJ, 879, 115 \\

\newpage

\appendix
\section{Numerical Algorithms for the Quantum Mechanical Approach}
\label{sec:appendix}

The quantum mechanical impact parameter probability of Vrinceanu \& Flannery (2001) 
can be written (Vrinceanu \etal 2012)
\beq
P_{ji}(R)=(2l'+1)\sum^{n-1}_{L=|l-l'|} (2L+1) 
\left\{\begin{array}{ccc} l' \quad l \quad L \\ j \quad j \quad j \end{array}\right\}^2
\label{PQM}
\\ \nonumber
\times \frac{(L!)^2(n-L-1)!}{(n+L)!}(2\sin\chi)^{2L} \left[ C^{(L+1)}_{n-L-1}(\cos\chi) \right]^2
\eeq
where $j=(n-1)/2$ and $C^{(\gamma)}_n$ denotes an ultraspherical (or Gegenbauer) 
polynomial.
The rotation angle $\chi$ between the orientation of the initial- and final-states is given by
\be
\cos\chi=\frac{1+\alpha^2\cos(\pi\sqrt{1+\alpha^2})}{1+\alpha^2}
\ee
for straight-line trajectories. The scattering parameter $\alpha$  is given by 
\be
\alpha=\frac{3Zn}{2vzR}
\ee
where $v$ denotes the speed of the collider.

Evaluation of both the ultraspherical polynomials and the $6j$-symbols $\left\{\ldots\right\}$ is numerically
challenging on considering principal quantum numbers up to $\sim 1000$ and for all allowed
orbital angular momenta due to under- \& over-flow and cancellation error. Vrinceanu \etal (2019) used high precision
(400 digits) to overcome this but note that it took 2 days of CPU time on
a single processor machine to evaluate the QM results of Figure \ref{fig3}.
We describe the algorithms that we use for their evaluation and which are many orders of
magnitude faster since they require only standard 64-bit floating point arithmetic for example.

\bigskip
\noindent{\it Quadrature:} We remark in passing that we evaluate all probability integrals using the trapezoidal rule
utilizing a logarithmic $\alpha$-mesh. This simultaneously yields both cross sections and rate coefficients.

\subsection{Ultraspherical Polynomials}
We describe a fast, accurate and stable algorithm for the evaluation of ultraspherical
polynomials $C^{(\gamma)}_{n}(x)$ as they occur in (\ref{PQM}).

We exploit the  fact that $\gamma+n$ is fixed in the summation.
Use Abramowitz \& Stegun (22.7.3):
\be
(n+1)C^{(\gamma)}_{n+1}(x)=2(n+\gamma)xC^{(\gamma)}_{n}(x)-(n+2\gamma-1)C^{(\gamma)}_{n-1}(x)
\label{UPC1}
\ee
to eliminate $C^{(\gamma)}_{n+1}$ from Abramowitz \& Stegun (22.7.23):
\be
(n+\gamma)C^{(\gamma-1)}_{n+1}(x)=(\gamma-1)\left[C^{(\gamma)}_{n+1}(x)-C^{(\gamma)}_{n-1}(x)\right]
\label{UPC2}
\ee
to obtain
\be
(n+1)C^{(\gamma-1)}_{n+1}(x)=2(\gamma-1)\left[xC^{(\gamma)}_{n}(x)-C^{(\gamma)}_{n-1}(x)\right]\,.
\label{UPC3}
\ee
Then use (\ref{UPC1}) again to eliminate $C^{(\gamma)}_{n-1}$  from (\ref{UPC2}) so as to obtain
\be
(n+2\gamma-1)C^{(\gamma)}_{n-1}(x)=2\gamma\left[C^{(\gamma+1)}_{n-1}(x)-xC^{(\gamma+1)}_{n-2}(x)\right]
\label{UPC4}
\ee
on relabelling $n\rightarrow n-2$ and $\gamma \rightarrow \gamma+1$.

Initialize $C^{(\gamma)}_{-1}(x)=0$ and $C^{(\gamma)}_{0}(x)=1$.
Then equations (\ref{UPC3}) and (\ref{UPC4}) can be used in tandem to make a single pass recurrence 
synchronized with the summation in (\ref{PQM}) which must start at the upper limit here.
The equations (\ref{UPC3}) and (\ref{UPC4}) are coupled directly here through the $C^{(\gamma)}_{n-1}(x)$ terms.

The above algorithm is applicable up to principal quantum number $n\approx 650$ using 64-bit
floating point arithmetic. Simply rescaling $C^{(\gamma)}_{0}(x)$ once extends the use of 64-bit arithmetic
up to $n\approx 1500$ without the need  to resort to higher precision. 
This is sufficiently high in $n$ so as to establish collisional LTE.

\subsection{Wigner \boldmath{$6j$}-symbols}
Racah (1942) first gave a closed expression for the recoupling of three angular momenta  to
give a resultant total --- the Racah W-coefficient --- which is written in terms of factorials.
These factorials can become rather large in practical applications and so subject to cancellation
error and under- \& over-flow when evaluated numerically. The Wigner $6j$-symbol is closely
related to the Racah W-coefficient but it exhibits the full symmetry of the problem (Edmonds 1957).

Consider the evaluation of the $6j$-symbol
\beq
\hspace{10mm}\left\{\begin{array}{ccc} a \quad b \quad c \\ d \quad e \quad f \end{array}\right\}\,.
\eeq
Define
\beq
w(j)=\left\{\begin{array}{ccc} j \quad b \quad c \\ d \quad e \quad f \end{array}\right\}
\eeq
where $b,c,d,e,f$ have been specified already.
Any $6j$-symbol can be re-ordered thus.
The $w(j)$ satisfy the following linear 3-term recurrence relation (Edmonds 1957, Schulten \& Gordon 1975a):
\be
jA(j+1)w(j+1)+B(j)w(j)+(j+1)A(j)w(j-1)=0
\label{WRR1}
\ee
for $j_\min\le j\le j_\max$ where $j_\min=\max\{|b-c|,|e-f|\}$ and $j_\max =\min\{|b+c|,|e+f|\}$.

The $A(j)$ and $B(j)$ correspond to specific $6j$-symbols with an argument $1/2$. They are given by
\beq
A(j)^2&=&[j^2-(b-c)^2][(b+c+1)^2-j^2]\nonumber\\
&\times&[j^2-(e-f)^2][(e+f+1)^2-j^2]
\eeq
and
\beq
B(j)/(2j+1)&=&j(j+1)[-j(j+1)+b(b+1)+c(c+1)]\nonumber\\
&+&e(e+1)[j(j+1)+b(b+1)-c(c+1)]\nonumber\\
&+&f(f+1)[j(j+1)-b(b+1)+c(c+1)]\nonumber\\
&-&2j(j+1)d(d+1)]\,.
\eeq
Note that $A(j_\min)=0=A(j_\max +1)$. A two-term relation then starts-off the unnormalized recurrence.
The solution is subsequently normalized via
\beq
\sum_j(2j+1)(2d+1)\left\{\begin{array}{ccc} j \quad b \quad c \\ d \quad e \quad f \end{array}\right\}^2=1\,.
\label{Wnorm}
\eeq
The phase is determined through
\beq
\mbox{sign} \left\{\begin{array}{ccc} j \quad b \quad c \\ d \quad e \quad f \end{array}\right\}=(-1)^{b+c+e+f}\,.
\eeq
The above linear 3-term recurrence relation (\ref{WRR1}) can be viewed as a finite-difference
relation for a second-order differential equation c.f. the Schr\"{o}dinger equation for a bound-state electron. 
It suffers a similar pathology to its solution.

We note that the range $j_\min \le j\le j_\max $ can be further sub-divided  as
\be
j_\min \le j\le j_\rmI \le j \le j_\rmII \le j \le j_\max
\ee
where $j_\rmI \le j \le j_\rmII$ defines the classically-allowed region of $w(j)$ and where the solution
is oscillatory as a function of $j$. This region corresponds to the resultant $j$ following the coupling 
of 3 angular momenta. 
These boundaries $j_\rmI$ and $j_\rmII$ (corresponding to the turning points $w''(j)=0$) 
can be determined from the root of a Cayley determinant (Schulten \& Gordon 1975b).

The required solution for $w(j)$ is exponentially decreasing 
in the classically-forbidden regions $j_\min \le j\le j_\rmI$ and $j_\rmII \le j \le j_\max $ 
as $j\rightarrow j_\min $ and $j\rightarrow j_\max $.
The recursion must then start at both ends and match somewhere in the classically-allowed region so as
to avoid picking-up the complementary exponentially increasing solution. Note that the use of a linear 3-term
recurrence relation in the classically-forbidden region leads to the need for constant rescaling so as to
avoid both numerical under- and over-flow.

The algorithm detailed above has been implemented by Schulten \& Gordon (1976) as the CPC program ACWQ.

The use of a non-linear 2-term recurrence relation
in the classically-forbidden region avoids the need for continual rescaling (Luscombe \& Luban 1998).
Define
%\vspace{-2mm}
\be
r(j)\equiv\frac{w(j)}{w(j-1)}\,.
\ee
Then the original recurrence relation (\ref{WRR1}) can be written as
\be
r(j)=\frac{-(j+1)A(j)}{B(j)+jA(j+1)r(j+1)}\quad\mbox{for}\quad j\le j_\max -1\,.
\ee
This defines a backwards recurrence with starting value 
%\vspace{-3mm}
\be
r(j_\max )=-(j_\max +1)A(j_\max )/B(j_\max )
%\vspace{-3mm}
\ee
since $A(j_\max +1)=0$.
Then $w(j)$ for $j_\rmII+1\le j \le j_\max $ is given by
\be
w(j_\rmII+k)=w(j_\rmII)\prod^k_{p=1}r(j_\rmII+p)
\label{WII}
\ee
for $1\le k\le j_\max -j_\rmII$. 
The value of $w(j_\rmII)$ at this point is both undefined and arbitrary.

This approach avoids under- and over-flow issues since $r(j)$ is bounded above by order unity. 
One can extend the evaluation somewhat into the classically-allowed region but must stop short 
of $w(j)$ changing sign so as to ensure that $w(j)\ne 0$.

Now define
\be
s(j)\equiv\frac{w(j)}{w(j+1)}\,.
\ee
Then the original recurrence relation (\ref{WRR1}) can be written as
\be
s(j)=\frac{-jA(j+1)}{B(j)+(j+1)A(j)s(j-1)}\quad\mbox{for}\quad j\ge j_\min +1\,.
\ee
This defines a forwards recurrence with starting value
\be
s(j_\min )=-j_\min A(j_\min +1)/B(j_\min )
\ee
since $A(j_\min )=0$.
Then $w(j)$ for $j_\min \le j \le j_\rmI-1$ is given by
\be
w(j_\rmI-k)=w(j_\rmI)\prod^k_{p=1}s(j_\rmI-p)
\label{WI}
\ee
for $1\le k\le j_\rmI- j_\min $. 
The value of $w(j_\rmI)$ is again both undefined and arbitrary.

We now need to determine $w(j)$ in the classically-allowed region and match with the
arbitrary/undefined $w(j_\rmI)$ and $w(j_\rmII)$. 
Define
\be
w_\rmI(j)\equiv\frac{w(j)}{w(j_\rmI)}\quad\mbox{and}\quad w_\rmII(j)\equiv\frac{w(j)}{w(j_\rmII)}\,.
\ee
These quantities $w_\rmI(j)$ and $w_\rmII(j)$ satisfy the original 3-term recurrence relation.
It is well-behaved in the classically-allowed region.

Use the initial values $w_\rmI(j_\rmI-1)=s(j_\rmI-1)$ and $w_\rmI(j_\rmI)=1$ so as to
carry-out a forwards recurrence for $w_\rmI(j)$ starting at $j=j_\rmI$ and on out to $j=j_m\le j_\rmII$.
Use the initial values $w_\rmII(j_\rmII+1)=r(j_\rmII+1)$ and $w_\rmII(j_\rmII)=1$ so as to
carry-out a backwards recurrence for $w_\rmII(j)$ starting at $j=j_\rmII$ and on in to $j=j_m\ge j_\rmI$.
Then we have that
\be
\frac{w_\rmII(j_m)}{w_\rmI(j_m)}=\frac{w(j_m)}{w(j_\rmII)}\times\frac{w(j_\rmI)}{w(j_m)}
=\frac{w(j_\rmI)}{w(j_\rmII)}\,.
\label{WII-I}
\ee
We see that our two unknowns $w(j_\rmI)$ and $w(j_\rmII)$ are reduced to a single unknown (ratio).

We have $w_\rmII(j)$ over $j_m\le j \le j_\rmII$. We obtain the remaining
values for $j_\rmI\le j \le j_m$ from
\be
w_\rmII(j)=w_\rmI(j)\times\frac{w(j_\rmI)}{w(j_\rmII)}\,.
\ee
We now have $w(j)$ over $j_\rmI\le j \le j_\rmII$:
\be
w(j)=w_\rmII(j)w(j_\rmII)
\ee
in terms of the unknown factor $w(j_\rmII)$. This factor can be determined 
through use of the normalization condition (\ref{Wnorm}). Then $w(j_\rmI)$  
can be determined from (\ref{WII-I}).
We already have $w(j)$ in the classically  forbidden region where it is 
 written in terms of $w(j_\rmI)$ and $w(j_\rmII)$ --- see (\ref{WII}) and (\ref{WI}).
This completes the determination of the $w(j)$.

The algorithm described above for the determination of $6j$-symbols is accurate for pathological cases such as
\beq
\left\{\begin{array}{ccc} 170/2 \quad 168/2 \quad 172/2 \\ 179/2 \quad 179/2 \quad 179/2 \end{array}\right\}
=3.3988213869\times 10^{-8}
\eeq
for which cancellation is an issue unless high precision is used.
There are no issues with regards to over/underflow. These  again require high
precision or constant re-scaling when using other algorithms.
We note that a $6j$-symbol with a value of $\lsrsim 10^{-16}$ is indistinguishable
from being identically zero in the classically-allowed region when using 64-bit floating point arithmetic.
We set such to zero.

We remark that this approach for $6j$-symbols can be adapted easily for the calculation of $3j$-symbols
as well (c.f. Schulten \& Gordon 1975a).

%The above algorithm has been implemented as a Fortran subroutine ({\sc wig6jr}) and deposited on Zenodo along with a simple interactive test driver.

\label{lastpage}
\clearpage
\end{document}